\documentclass[aps,manuscript,pop,showpacs,preprint,superscriptaddress]{revtex4-1}
\usepackage{lmodern}

\usepackage{lmodern}
\usepackage[T1]{fontenc}
\usepackage[latin9]{inputenc}
\usepackage{color}
\usepackage{amsmath}
\usepackage{graphicx}

\makeatletter

\providecommand{\tabularnewline}{\\}

\@ifundefined{textcolor}{}
{%
 \definecolor{BLACK}{gray}{0}
 \definecolor{WHITE}{gray}{1}
 \definecolor{RED}{rgb}{1,0,0}
 \definecolor{GREEN}{rgb}{0,1,0}
 \definecolor{BLUE}{rgb}{0,0,1}
 \definecolor{CYAN}{cmyk}{1,0,0,0}
 \definecolor{MAGENTA}{cmyk}{0,1,0,0}
 \definecolor{YELLOW}{cmyk}{0,0,1,0}
}

\usepackage{amsthm}\usepackage{latexsym}\usepackage{bm}\usepackage{amsfonts}

\makeatother

\begin{document}

\title{Explicit symplectic algorithms based on generating functions for
charged particle dynamics }

\author{Ruili Zhang}

\affiliation{Department of Modern Physics and School of Nuclear Science and Technology,
University of Science and Technology of China, Hefei, Anhui 230026,
China}

\affiliation{Key Laboratory of Geospace Environment, CAS, Hefei, Anhui 230026,
China}

\author{Hong Qin }

\thanks{Corresponding author. hongqin@ustc.edu.cn}

\affiliation{Department of Modern Physics and School of Nuclear Science and Technology,
University of Science and Technology of China, Hefei, Anhui 230026,
China}

\affiliation{Plasma Physics Laboratory, Princeton University, Princeton, NJ 08543,
USA}

\author{Yifa Tang}

\affiliation{LSEC, Academy of Mathematics and Systems Science, Chinese Academy
of Sciences, Beijing 100190, China}

\author{Jian Liu}

\affiliation{Department of Modern Physics and School of Nuclear Science and Technology,
University of Science and Technology of China, Hefei, Anhui 230026,
China}

\affiliation{Key Laboratory of Geospace Environment, CAS, Hefei, Anhui 230026,
China}

\author{Yang He}

\affiliation{Department of Modern Physics and School of Nuclear Science and Technology,
University of Science and Technology of China, Hefei, Anhui 230026,
China}

\affiliation{Key Laboratory of Geospace Environment, CAS, Hefei, Anhui 230026,
China}

\author{Jianyuan Xiao}

\affiliation{Department of Modern Physics and School of Nuclear Science and Technology,
University of Science and Technology of China, Hefei, Anhui 230026,
China}

\affiliation{Key Laboratory of Geospace Environment, CAS, Hefei, Anhui 230026,
China}
\begin{abstract}
Dynamics of a charged particle in the canonical coordinates is a Hamiltonian
system, and the well-known symplectic algorithm has been regarded
as the \emph{de facto} method for numerical integration of Hamiltonian
systems due to its long-term accuracy and fidelity. For long-term
simulations with high efficiency, explicit symplectic algorithms are
desirable. However, it is widely accepted that explicit symplectic
algorithms are only available for sum-separable Hamiltonians, and
that this restriction severely limits the application of explicit
symplectic algorithms to charged particle dynamics. To overcome this
difficulty, we combine the familiar sum-split method and a generating
function method to construct second and third order explicit symplectic
algorithms for dynamics of charged particle. The generating function
method is designed to generate explicit symplectic algorithms for
product-separable Hamiltonian with form of $H(\mathbf{p},\mathbf{q})=\mathbf{p}_{i}f(\mathbf{q})$
or $H(\mathbf{p},\mathbf{q})=\mathbf{q}_{i}f(\mathbf{p})$. Applied
to the simulations of charged particle dynamics, the explicit symplectic
algorithms based on generating functions demonstrate superiorities
in conservation and efficiency.
\end{abstract}
\maketitle

\section{Introduction}

The dynamics of a charged particle with the Lorentz force in the canonical
coordinates $(\mathbf{x},\mathbf{p})$ is a canonical Hamiltonian
system,
\begin{equation}
\dfrac{d\mathbf{Z}}{dt}=J^{-1}\nabla H(\mathbf{Z}):=\begin{cases}
 & \begin{split} & \dfrac{d\mathbf{x}}{dt}=\dfrac{1}{m}\left(\mathbf{p}-q\mathbf{A}(\mathbf{x})\right)\,,\\
 & \dfrac{d\mathbf{p}}{dt}=-q\nabla\phi(\mathbf{x})+\dfrac{q}{m}\left(\dfrac{\partial\mathbf{A}}{\partial\mathbf{x}}\right)^{T}(\mathbf{p-}q\mathbf{A})\,,
\end{split}
\end{cases}\label{eq:nr}
\end{equation}
where $\mathbf{Z}=(\mathbf{x}^{T},\mathbf{p}^{T})^{T}$ is a 6-dimensional
vector,
\[
\thinspace J=\left(\begin{array}{cc}
0 & -I\\
I & 0
\end{array}\right)\thinspace
\]
is the canonical symplectic matrix and
\begin{equation}
H(\mathbf{Z})=\dfrac{1}{2m}(\mathbf{p}-q\mathbf{A}(\mathbf{x}))^{2}+q\phi(\mathbf{x})\label{eq:Hnr}
\end{equation}
is the Hamiltonian function. For canonical Hamiltonian system
\begin{align}
\dot{\mathbf{Z}} & =J^{-1}\nabla H(\mathbf{Z})\thinspace,\mathbf{\; Z}\in\mathbf{R}^{2k}\,,\;\mathbf{Z}(t_{0})=\mathbf{Z}_{0},\label{eq:Ham}
\end{align}
it is well known that symplectic algorithms conserve the symplectic
structure exactly and globally bound the energy error by a small number
\cite{ruth1983canonical,feng1985,feng1986,kang1989construction,forest1989fourth,channell1990symplectic,yoshida1990construction,candy1991symplectic,SSC94,feng1995collected,yoshida1993recent,marsden2001discrete,GNT,feng2010symplectic}.
They have become the\emph{ de facto} standard for numerical integration
of Hamiltonian systems with important applications in nonlinear dynamics,
astrophysics, plasma physics, accelerator physics, and quantum physics.
Recently, symplectic and geometric algorithms have been developed
for non-canonical particle dynamics \textcolor{black}{\cite{qin2008variational,qin2009variational,guan2010phase,squire2012gauge,qin2013boris,liu2014fate,zhang2014,zhang2015volume,ellison2015comment,he2015volume,he2015explicit,ellison2015development,liu2015neoclassical,He16-172}}
and the infinite dimensional particle-field systems \textcolor{black}{\cite{Squire4748,squire2012geometric,xiao2013variational,kraus2013variational,evstatiev2013variational,zhou2014variational,Shadwick14,xiao2015variational,xiao2015explicit,crouseilles2015hamiltonian,qin2015comment,he2015Hamiltonian,qin2016canonical,zhou2015formation,Webb16}}
in plasma physics and accelerator physics. To improve the efficiency
and accuracy of long-term simulations for systems with a large number,
e.g., $10^{9}$, of degrees of freedom, explicit symplectic algorithms
are desired. Strictly speaking, in order for implicit symplectic algorithms
to be symplectic, the implicit iteration relations need to be solved
exactly, which is impossible in general. The best one can expect is
to solve the implicit iteration relations to machine precision at
an extreme cost. This is the main reason to search for explicit symplectic
method. Splitting method has been proven to be an effective tool in
constructing explicit symplectic algorithms \cite{mclachlan2002splitting,sheng1989solving,chin2007higher,chin2008symplectic,chin2010multi,he2015volume,zhang2015volume,crouseilles2015hamiltonian,qin2015comment}.
The basic procedure is to decompose original system into solvable
subsystems possessing the same geometric structure, then compose the
geometric sub-algorithms together to obtain the desired algorithms
\cite{mclachlan2002splitting,mclachlan2002families,GNT}. It is well
known that for a Hamiltonian whose $\mathrm{\mathbf{p}}$-dependence
and $\mathbf{q}$-dependence can be separated as sumands in a summation
as follows
\begin{equation}
H(\mathbf{Z})=f(\mathbf{p})+g(\mathbf{q})\thinspace,\label{eq:HS}
\end{equation}
the splitting method can generate explicit symplectic algorithms of
any orders \cite{yoshida1990construction,forest1989fourth}. The familiar
leapfrog algorithm is an example of this method. We will call the
form in Eq.\,\eqref{eq:HS} \emph{sum-separable} and refer to this
well-known splitting method \emph{as sum-split method}. It is generally
believed that if a Hamiltonian is not sum-separable as in Eq.\,\eqref{eq:HS},
general explicit symplectic algorithms do not exist\textbf{\textcolor{red}{{}
}}\textcolor{black}{\cite{yoshida1993recent,mclachlan2002splitting,chin2008symplectic,chin2009explicit,feng2010symplectic,blanes2012explicit,GNT}}\emph{.}
For dynamics of charged particle, sum-split method loses efficacy
and can not be applied directly to construct explicit symplectic algorithms,
because the Hamiltonian Eq.\,\eqref{eq:Hnr} is not sum-separable.
An explicit non-canonical symplectic algorithm has been developed
by He et al. using sum-split method for charged particle dynamics
in the non-canonical coordinates \cite{he2015explicit,he2015Hamiltonian,xiao2015explicit}.
However, it requires numerical integration of the magnetic field along
given paths, which can be non-trivial for certain complicated magnetic
fields. In this paper, different from He's splitting algorithm, we
combine the familiar sum-split method with a generating function method
to construct explicit symplectic algorithms for dynamics of charged
particles, which do not require numerical integration of the magnetic
field.

The generating function method has been well developed to construct
symplectic methods for a Hamiltonian system Eq.\,\eqref{eq:Ham}
\cite{kang1989construction,GNT,feng2010symplectic}. There are three
types generating functions utilized to construct different types of
symplectic algorithms. The symplectic Euler method and mid-point method
are included in this family. Generally speaking, symplectic methods
based on all three types of generating functions are usually implicit.
However, for\emph{ product-separable} Hamiltonians in the form of
\begin{equation}
H(\mathbf{Z})=\mathbf{p}_{i}g(\mathbf{q})\thinspace,\label{eq:HPS1}
\end{equation}
or
\begin{equation}
H(\mathbf{Z})=\mathbf{q}_{i}g(\mathbf{p})\thinspace,\label{eq:HPS2}
\end{equation}
explicit symplectic algorithms with accuracy of order 2 and 3 can
be constructed by applying the first type generating function and
the second type generating function respectively. Here, \emph{product-separable}
means that the $\mathbf{q}$-dependency and \textbf{$\mathbf{p}$}-dependency
can be separated as factors in a production. For dynamics of charged
particle governed by Eq.\,\eqref{eq:nr}, we sum-split the Hamiltonian
Eq.\,\eqref{eq:Hnr} into five parts, two of which can be solved
exactly. The other three parts are in the form of Eq.\,\eqref{eq:HPS1},
and admit explicit symplectic algorithms based on the generating functions.
Then combining the exact solution flows and explicit symplectic sub-algorithms
in various manners, explicit symplectic algorithms of different orders
can be constructed.

The paper is organized as follows. In Sec.$\,$II, symplectic algorithms
based on generating functions are introduced, and for the Hamiltonian
systems with the forms of Eqs.\,\eqref{eq:HPS1} and \eqref{eq:HPS2},
explicit symplectic algorithms are given. In Sec.\,III, we construct
explicit symplectic algorithms of order 2 and 3 for charged particle
dynamics based on generating functions. Numerical experiments are
provided, and the superiority of the explicit symplectic algorithms
relative to non-symplectic Runge-Kutta methods and implicit symplectic
methods is demonstrated in Sec.\,IV.

\section{Symplectic method based on generating function}

For a Hamiltonian system Eq.\,\eqref{eq:Ham}, we introduce symplectic
methods based on the first and second type of generating functions.
The symplectic methods based on generating functions of the first
type can be written as
\begin{equation}
\begin{cases}
\mathbf{p}^{n+1} & =\mathbf{p}^{n}-\nabla_{q}G(\mathbf{p}^{n+1},\mathbf{q}^{n},\Delta t)\thinspace,\\
\mathbf{q}^{n+1} & =\mathbf{q}^{n}+\nabla_{p}G(\mathbf{p}^{n+1},\mathbf{q}^{n},\Delta t)\thinspace,
\end{cases}
\end{equation}
with the generating function
\begin{equation}
G(\mathbf{p},\mathbf{q},t)=tG_{1}(\mathbf{p},\mathbf{q})+t^{2}G_{2}(\mathbf{p},\mathbf{q})+t^{3}G_{3}(\mathbf{p},\mathbf{q})+\cdots\thinspace,
\end{equation}
where
\begin{equation}
\begin{split}G_{1}(\mathbf{p},\mathbf{q}) & =H(\mathbf{p},\mathbf{q})\thinspace,\\
G_{2}(\mathbf{p},\mathbf{q}) & =\dfrac{1}{2}\left(\dfrac{\partial H}{\partial\mathbf{p}}\dfrac{\partial H}{\partial\mathbf{q}}\right)(\mathbf{p},\mathbf{q})\thinspace,\\
G_{3}(\mathbf{p},\mathbf{q}) & =\dfrac{1}{6}\left[\dfrac{\partial^{2}H}{\partial\mathbf{p}^{2}}\left(\dfrac{\partial H}{\partial\mathbf{q}}\right)^{2}+\dfrac{\partial^{2}H}{\partial\mathbf{p}\partial\mathbf{q}}\dfrac{\partial H}{\partial\mathbf{p}}\dfrac{\partial H}{\partial\mathbf{q}}+\dfrac{\partial^{2}H}{\partial\mathbf{q}^{2}}\left(\dfrac{\partial H}{\partial\mathbf{p}}\right)^{2}\right]\thinspace,\\
......\thinspace\text{.}
\end{split}
\end{equation}
Utilizing the truncated series,
\begin{equation}
G(\mathbf{p},\mathbf{q},t)=\sum_{i=1}^{r}t^{i}G_{i}(\mathbf{p},\mathbf{q})\thinspace.
\end{equation}
we obtain a symplectic method of order $r$ \cite{kang1989construction,GNT,feng2010symplectic}.
The symplectic methods based on generating functions of the second
type can be constructed similarly. Both types are usually implicit
for general Hamiltonian systems. However, for product-separable Hamiltonian
with the form of Eq.\,\eqref{eq:HPS1} or Eq.\,\eqref{eq:HPS2},
second and third order symplectic algorithms based on generating functions
can be constructed explicitly. Let's take Hamiltonian Eq.\,\eqref{eq:HPS1}
as an example to demonstrate the explicit symplectic methods based
on the generating functions. The corresponding second order generating
function of type one is
\begin{equation}
G(\mathbf{p},\mathbf{q},t)=t\mathbf{p}_{i}f(\mathbf{q})+\dfrac{t^{2}}{2}\mathbf{p}_{i}\dfrac{\partial f}{\partial q_{i}}f(\mathbf{q})\thinspace,
\end{equation}
Then the explicit symplectic method of order 2 based on the generating
function is
\begin{equation}
\begin{cases}
\mathbf{p}^{n+1} & =\mathbf{p}^{n}-\mathbf{p}_{i}^{n+1}\left[\Delta t\nabla_{q}f(\mathbf{q})+\dfrac{\Delta t^{2}}{2}\nabla_{q}\left(\dfrac{\partial f}{\partial q_{i}}f(\mathbf{q}^{n})\right)\right]\thinspace,\\
\mathbf{q}_{i}^{n+1} & =\mathbf{q}_{i}^{n}+\Delta tf(\mathbf{q}^{n})+\dfrac{\Delta t^{2}}{2}\dfrac{\partial f}{\partial q_{i}}f(\mathbf{q}^{n})\thinspace.
\end{cases}
\end{equation}
For the product-separable Hamiltonian in the form of Eq.\,\eqref{eq:HPS2},
explicit symplectic algorithms can be constructed similarly utilizing
generating functions of the second type.

\section{Explicit symplectic algorithms for charged particle dynamics}

\begin{figure}
\includegraphics[scale=0.55]{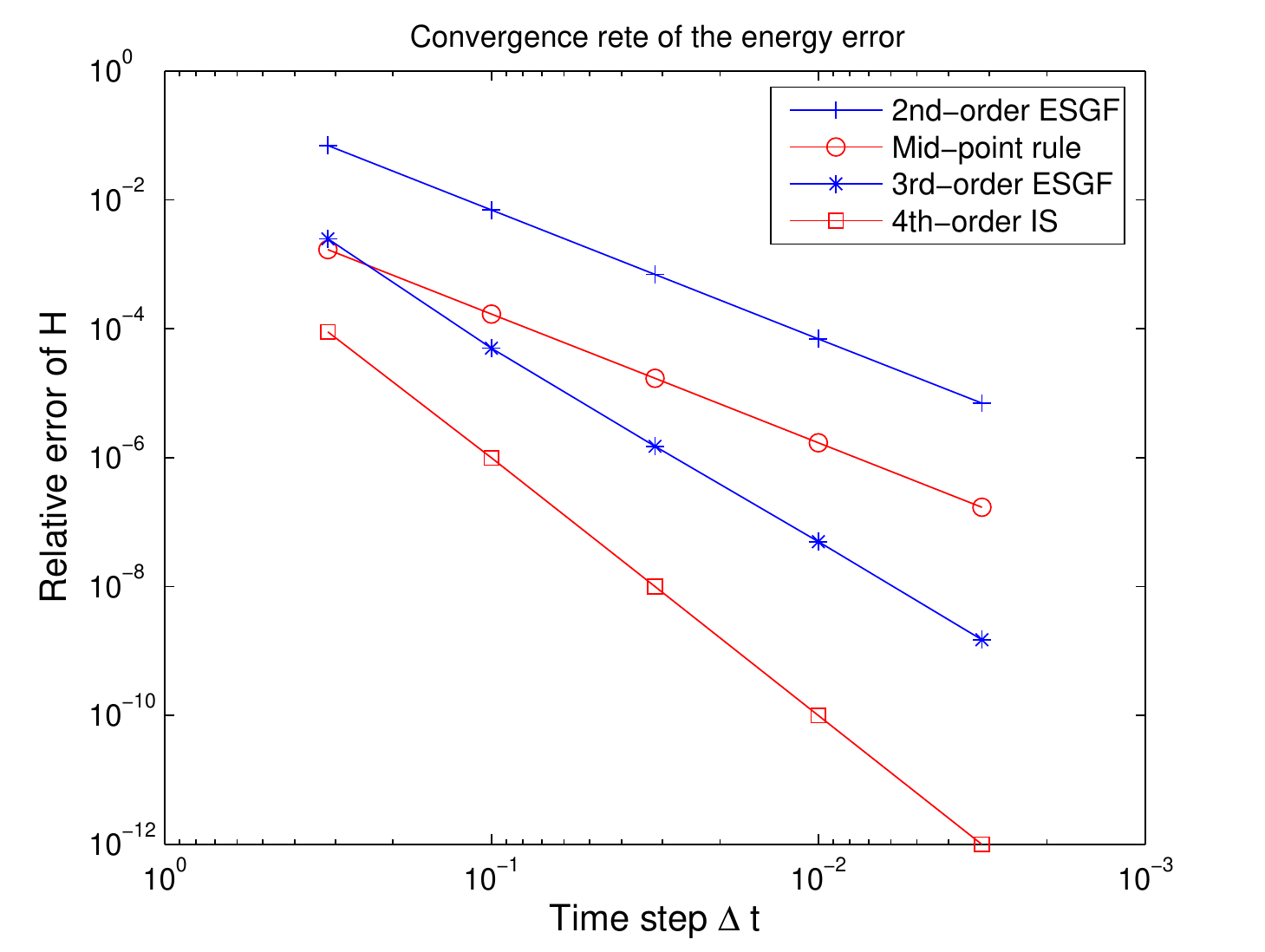}

\protect\caption{Convergence rate of the energy error for four symplectic methods.
It verifies that $\Psi_{\Delta t}^{2}$ is indeed a second order method
and $\Psi_{\Delta t}^{3}$ is a third order method.\label{fig1}}
\end{figure}

In this section, we will use the methods given in Sec.\,II to construct
explicit symplectic algorithms for charged particle dynamics determined
by Eq.\,\eqref{eq:nr}. It was commonly believed that this system
does not admit any explicit symplectic algorithm, because the Hamiltonian
given by Eq.\,\eqref{eq:Hnr} is not sum-separable. Now, we show
how to construct explicit symplectic algorithms for it using the generating-function
method and the familiar sum-split method. We sum-split the Hamiltonian
function into five parts as
\begin{equation}
H(\mathbf{x},\mathbf{p})=H_{1}+H_{2}+H_{3}+H_{4}+H_{5}\,,
\end{equation}
where
\begin{equation}
\begin{split} & H_{1}=\dfrac{1}{2m}\mathbf{p}^{2}\,,\qquad H_{2}=\dfrac{q^{2}}{2m}\mathbf{A}(\mathbf{x})^{2}+q\phi(\mathbf{x})\,,\\
 & H_{3}=-\dfrac{q}{m}\mathbf{A}(\mathbf{x})^{T}\left(p_{1},0,0\right)^{T}=-\dfrac{q}{m}\mathbf{A}_{1}(\mathbf{x})p_{1}\,,\\
 & H_{4}=-\dfrac{q}{m}\mathbf{A}(\mathbf{x})^{T}\left(0,p_{2},0\right)^{T}=-\dfrac{q}{m}\mathbf{A}_{2}(\mathbf{x})p_{2}\,,\\
 & H_{5}=-\dfrac{q}{m}\mathbf{A}(\mathbf{x})^{T}\left(0,0,p_{3}\right)^{T}=-\dfrac{q}{m}\mathbf{A}_{3}(\mathbf{x})p_{3\,.}
\end{split}
\end{equation}
The corresponding sub-systems generated by these sub-Hamiltonians
are
\begin{equation}
\begin{split} & S_{1}:=\begin{cases}
 & \dfrac{d\mathbf{x}}{dt}=\dfrac{1}{m}\mathbf{p}\,,\\
 & \dfrac{d\mathbf{p}}{dt}=\mathbf{0}\,,
\end{cases}\\
 & S_{2}:=\begin{cases}
 & \dfrac{d\mathbf{x}}{dt}=\mathbf{0}\,,\\
 & \dfrac{d\mathbf{p}}{dt}=-\dfrac{q^{2}}{m}\left(\dfrac{\partial\mathbf{A}}{\partial\mathbf{x}}\right)^{T}\mathbf{A}-q\nabla\phi(\mathbf{x})\,,
\end{cases}\\
 & S_{3}:=\begin{cases}
 & \dfrac{d\mathbf{x}}{dt}=\mathbf{-}\dfrac{q}{m}(\mathbf{A}_{1}(\mathbf{x}),0,0)^{T}\,,\\
 & \dfrac{d\mathbf{p}}{dt}=\dfrac{q}{m}\left(\dfrac{\partial\mathbf{A}}{\partial\mathbf{x}}\right)^{T}\left(p_{1},0,0\right)^{T}\,,
\end{cases}\\
 & S_{4}:=\begin{cases}
 & \dfrac{d\mathbf{x}}{dt}=\mathbf{-}\dfrac{q}{m}(0,\mathbf{A}_{2}(\mathbf{x}),0)^{T}\,,\\
 & \dfrac{d\mathbf{p}}{dt}=\dfrac{q}{m}\left(\dfrac{\partial\mathbf{A}}{\partial\mathbf{x}}\right)^{T}\left(0,p_{2},0\right)^{T}\,,
\end{cases}\:\\
 & S_{5}:=\begin{cases}
 & \dfrac{d\mathbf{x}}{dt}=\mathbf{-}\dfrac{q}{m}(0,0,\mathbf{A}_{3}(\mathbf{x}),)^{T}\,,\\
 & \dfrac{d\mathbf{p}}{dt}=\dfrac{q}{m}\left(\dfrac{\partial\mathbf{A}}{\partial\mathbf{x}}\right)^{T}\left(0,0,p_{3}\right)^{T}\,.
\end{cases}
\end{split}
\end{equation}
For subsystems $S_{1}$ and $S_{2}$, exact solutions can be computed
explicitly as
\begin{align}
\begin{split} & \varphi_{1}(t):=\begin{cases}
 & \mathbf{x}(t)=\mathbf{x}_{0}+t\dfrac{1}{m}\mathbf{p}_{0}\,,\\
 & \mathbf{p}(t)=\mathbf{p}_{0}\,,
\end{cases}\:\\
 & \varphi_{2}(t):=\begin{cases}
 & \mathbf{x}(t)=\mathbf{x}_{0}\,,\\
 & \mathbf{p}(t)=\mathbf{p}_{0}-t\dfrac{q^{2}}{m}\left(\dfrac{\partial\mathbf{A}}{\partial\mathbf{x}}\right)^{T}\mathbf{A}\mid_{\mathbf{x}=\mathbf{x}_{0}}-qt\nabla\phi(\mathbf{x}_{0})\,.
\end{cases}
\end{split}
\end{align}
The sub-Hamiltonians of remaining three subsystems $S_{3}$, $S_{4}$
and $S_{5}$ are all product-separable as in Eq.\,\eqref{eq:HPS1}.
Let's take the sub-system $S_{3}$ associated with the sub-Hamiltonian
$H_{3}(\mathbf{p},\mathbf{x})=-\dfrac{q}{m}p_{1}\mathbf{A}_{1}(\mathbf{x})$
as an example to demonstrate our method. In terms of Cartesian components,
the sub-system $S_{3}$ is
\begin{equation}
S_{3}:=\begin{cases}
 & \dfrac{dx}{dt}=-\dfrac{q}{m}\mathbf{A}_{1}(\mathbf{x})\,,\\
 & \dfrac{dp_{1}}{dt}=\dfrac{q}{m}\dfrac{\partial\mathbf{A}_{1}}{\partial x}p_{1}\,,\\
 & \dfrac{dp_{2}}{dt}=\dfrac{q}{m}\dfrac{\partial\mathbf{A}_{1}}{\partial y}p_{1}\,,\\
 & \dfrac{dp_{3}}{dt}=\dfrac{q}{m}\dfrac{\partial\mathbf{A}_{1}}{\partial z}p_{1}\,.
\end{cases}\label{eq:S3}
\end{equation}
The symplectic method of order 2 based on generating function can
be obtained,
\begin{equation}
\begin{cases}
\mathbf{p}^{n+1} & =\mathbf{p}^{n}-\nabla_{\mathbf{x}}G(\mathbf{p}^{n+1},\mathbf{x}^{n},\Delta t)\thinspace,\\
\mathbf{x}^{n+1} & =\mathbf{x}^{n}+\nabla_{\mathbf{p}}G(\mathbf{p}^{n+1},\mathbf{x}^{n},\Delta t)\thinspace,
\end{cases}
\end{equation}
where the truncated generating function of order 2 is
\begin{equation}
\begin{split}G(\mathbf{p},\mathbf{x},\Delta t) & =\Delta tH_{3}(\mathbf{p},\mathbf{x})+\dfrac{\Delta t^{2}}{2}\left(\nabla_{\mathbf{p}}H_{3}\cdot\nabla_{\mathbf{x}}H_{3}\right)(\mathbf{p},\mathbf{x})\thinspace,\\
 & =-\Delta t\dfrac{q}{m}p_{1}\mathbf{A}_{1}(\mathbf{x})+\dfrac{\Delta t^{2}}{2}\dfrac{q^{2}}{m^{2}}p_{1}\dfrac{\partial\mathbf{A}_{1}}{\partial x}\mathbf{A}_{1}(\mathbf{x})\thinspace.
\end{split}
\end{equation}
Thus, the second-order symplectic methods for $S_{3}$ is
\begin{equation}
\psi_{3}^{\Delta t}=\begin{cases}
x^{n+1}=x^{n}-\Delta t\dfrac{q}{m}\mathbf{A}_{1}(x^{n},y^{n},z^{n})+\dfrac{\Delta t^{2}}{2}\dfrac{q^{2}}{m^{2}}\mathbf{A}_{1}(x^{n},y^{n},z^{n})\dfrac{\partial\mathbf{A}_{1}}{\partial x}(x^{n},y^{n},z^{n})\,,\\
p_{1}^{n+1}=p_{1}^{n}+p_{1}^{n+1}\left[\Delta t\dfrac{q}{m}\dfrac{\partial\mathbf{A}_{1}}{\partial x}-\dfrac{\Delta t^{2}}{2}\dfrac{q^{2}}{m^{2}}\dfrac{\partial\mathbf{A}_{1}}{\partial x}\dfrac{\partial\mathbf{A}_{1}}{\partial x}-\dfrac{\Delta t^{2}}{2}\dfrac{q^{2}}{m^{2}}\mathbf{A}_{1}\dfrac{\partial^{2}\mathbf{A}_{1}}{\partial x\partial x}\right](x^{n},y^{n},z^{n})\,,\\
p_{2}^{n+1}=p_{2}^{n}+p_{1}^{n+1}\left[\Delta t\dfrac{q}{m}\dfrac{\partial\mathbf{A}_{1}}{\partial y}-\dfrac{\Delta t^{2}}{2}\dfrac{q^{2}}{m^{2}}\dfrac{\partial\mathbf{A}_{1}}{\partial x}\dfrac{\partial\mathbf{A}_{1}}{\partial y}-\dfrac{\Delta t^{2}}{2}\dfrac{q^{2}}{m^{2}}\mathbf{A}_{1}\dfrac{\partial^{2}\mathbf{A}_{1}}{\partial x\partial y}\right](x^{n},y^{n},z^{n})\thinspace,\\
p_{3}^{n+1}=p_{3}^{n}+p_{1}^{n+1}\left[\Delta t\dfrac{q}{m}\dfrac{\partial\mathbf{A}_{1}}{\partial z}-\dfrac{\Delta t^{2}}{2}\dfrac{q^{2}}{m^{2}}\dfrac{\partial\mathbf{A}_{1}}{\partial x}\dfrac{\partial\mathbf{A}_{1}}{\partial z}-\dfrac{\Delta t^{2}}{2}\dfrac{q^{2}}{m^{2}}\mathbf{A}_{1}\dfrac{\partial^{2}\mathbf{A}_{1}}{\partial x\partial z}\right](x^{n},y^{n},z^{n})\thinspace,
\end{cases}
\end{equation}
which is an explicit method, but not symmetric. For sub-systems $S_{4},$
and $S_{5}$, second order explicit symplectic methods $\psi_{4}^{\Delta t}$
and $\psi_{5}^{\Delta t}$ are constructed similarly. Composing the
exact solutions and the symplectic numerical flows of the five subsystems,
we obtain the following explicit symplectic method for charged particle
dynamics with the accuracy of order 1,
\begin{equation}
\Psi_{\Delta t}^{1}=\varphi_{1}^{\Delta t}\circ\varphi_{2}^{\Delta t}\circ\psi_{3}^{\Delta t}\circ\psi_{4}^{\Delta t}\circ\psi_{5}^{\Delta t}\,.
\end{equation}
If the sub-numerical solution $\psi_{3}^{\Delta t}$, $\psi_{4}^{\Delta t}$,
and $\psi_{5}^{\Delta t}$ were symmetric, the symplectic method obtained
by symmetric composition
\begin{equation}
\Psi_{\Delta t}^{2}=\varphi_{1}^{\Delta t/2}\circ\varphi_{2}^{\Delta t/2}\circ\psi_{3}^{\Delta t/2}\circ\psi_{4}^{\Delta t/2}\circ\psi_{5}^{\Delta t}\circ\psi_{4}^{\Delta t/2}\circ\psi_{3}^{\Delta t/2}\circ\varphi_{2}^{\Delta t/2}\circ\varphi_{1}^{\Delta t/2}\,\label{eq:2thes}
\end{equation}
would be symmetric and of order 2. Because $\psi_{3}^{\Delta t}$,
$\psi_{4}^{\Delta t}$, and $\psi_{5}^{\Delta t}$ are not symmetric,
neither is $\Psi_{\Delta t}^{2}$. However, we can prove that $\Psi_{\Delta t}^{2}$
is of second order. The proof is given in the Appendix. Since all
the sub-algorithms preserve the canonical symplectic structure, $\Psi_{\Delta t}^{1}$
and $\Psi_{\Delta t}^{2}$ preserve the canonical symplectic structure
naturally. Of course, it is possible to increase the accuracy of the
numerical methods by various compositions \cite{mclachlan2002splitting,GNT}.
For example, a third order algorithm can be obtained by the following
composition method using $\Psi_{\Delta t}^{2}$,
\begin{equation}
\Psi_{\Delta t}^{3}=\Psi_{a\Delta t}^{2}\circ\Psi_{b\Delta t}^{2}\circ\Psi_{a\Delta t}^{2}\thinspace,
\end{equation}
where $a=\dfrac{1}{2-2^{1/3}}$ and $b=1-2a$. To numerically verify
the orders of $\Psi_{\Delta t}^{2}$ and $\Psi_{\Delta t}^{3}$, we
now apply $\Psi_{\Delta t}^{2}$, $\Psi_{\Delta t}^{3}$, the second
order implicit mid-point method and a 4th-order implicit symplectic
method to simulate the dynamics of charged particle in the magnetic
field of a tokamak (see next section). Here, the 4th-order implicit
symplectic method is generated by symmetric composition of the second
order implicit mid-point method. The relative errors of Hamiltonian
as functions of time step $\Delta t$ for these methods are plotted
in Fig. \ref{fig1}, which verifies that $\Psi_{\Delta t}^{2}$ is
indeed a second order method and $\Psi_{\Delta t}^{3}$ is a third
order method.

\section{Numerical examples}

To numerically test the explicit symplectic algorithms developed,
we simulate the dynamics of a $3.5$MeV $\alpha-$particle, which
is a product of D-T fusion, in the magnetic field of a tokamak. We
will compare the second order explicit symplectic (ES2) method $\Psi_{\Delta t}^{2}$
developed with the second order implicit symplectic mid-point (IS2)
method and the third order non-symplectic Runge-Kutta (RK3) method.
Numerical results will demonstrate the superb properties of explicit
symplectic methods in terms of accuracy, efficiency and preserving
energy over long-term simulations.

\begin{figure}
\includegraphics[scale=0.8]{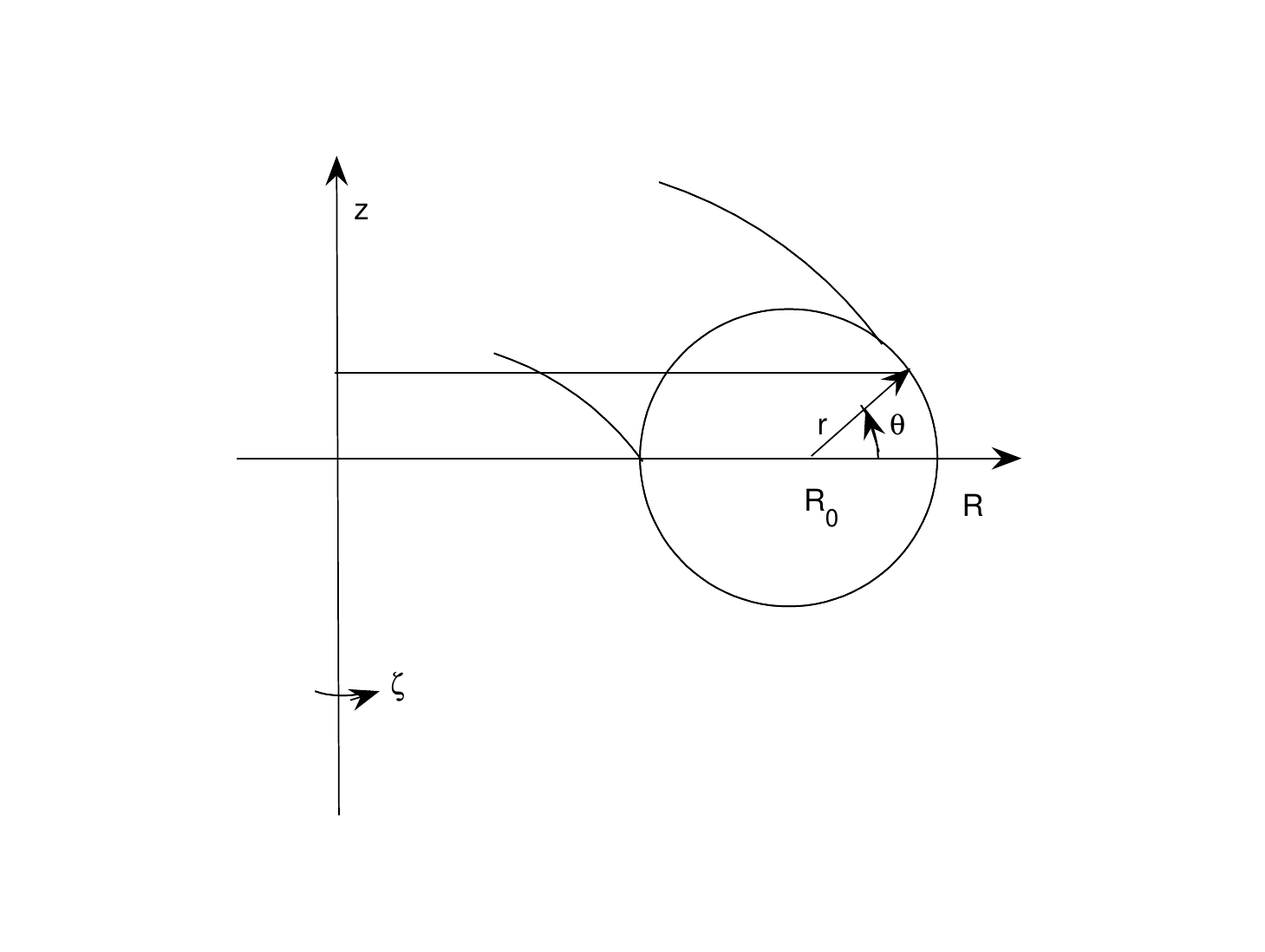}

\protect\caption{2D tokamak geometry with circular concentric flux surfaces. \label{fig2}}
\end{figure}

The axisymmetric tokamak geometry is illustrated in \textcolor{black}{Fig.
\ref{fig2}. }A model vector potential of the magnetic field is
\begin{equation}
\mathbf{A}=\dfrac{B_{0}r^{2}}{2Rq}e_{\zeta}-\ln\left(\dfrac{R}{R_{0}}\right)\dfrac{R_{0}B_{0}}{2}e_{z}+\dfrac{B_{0}R_{0}z}{2R}e_{R}\,,
\end{equation}
where $R=\sqrt{x^{2}+y^{2}}$ is the major radius coordinate, $R_{0}$
is the major radius, $B_{0}$ is the magnetic field on axis, the constant
$q$ is the safety factor, and $\zeta=arctan\left(\frac{x}{y}\right)$
is the toroidal coordinate of the torus. In this example, we take
$R_{0}=3m$ and $B_{0}=1T$ with $q=2$.

\begin{figure}
\includegraphics[scale=0.55]{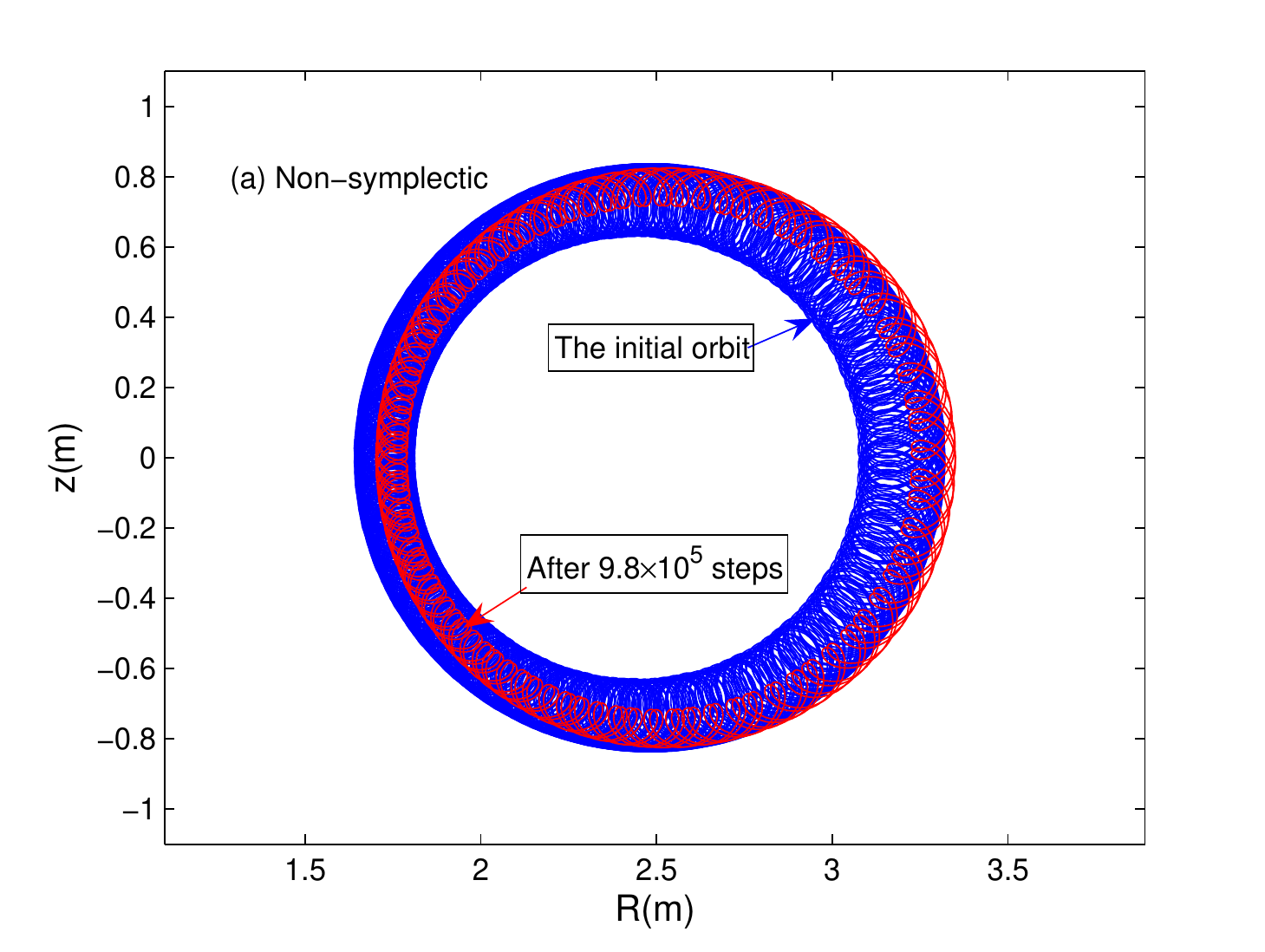}\includegraphics[scale=0.55]{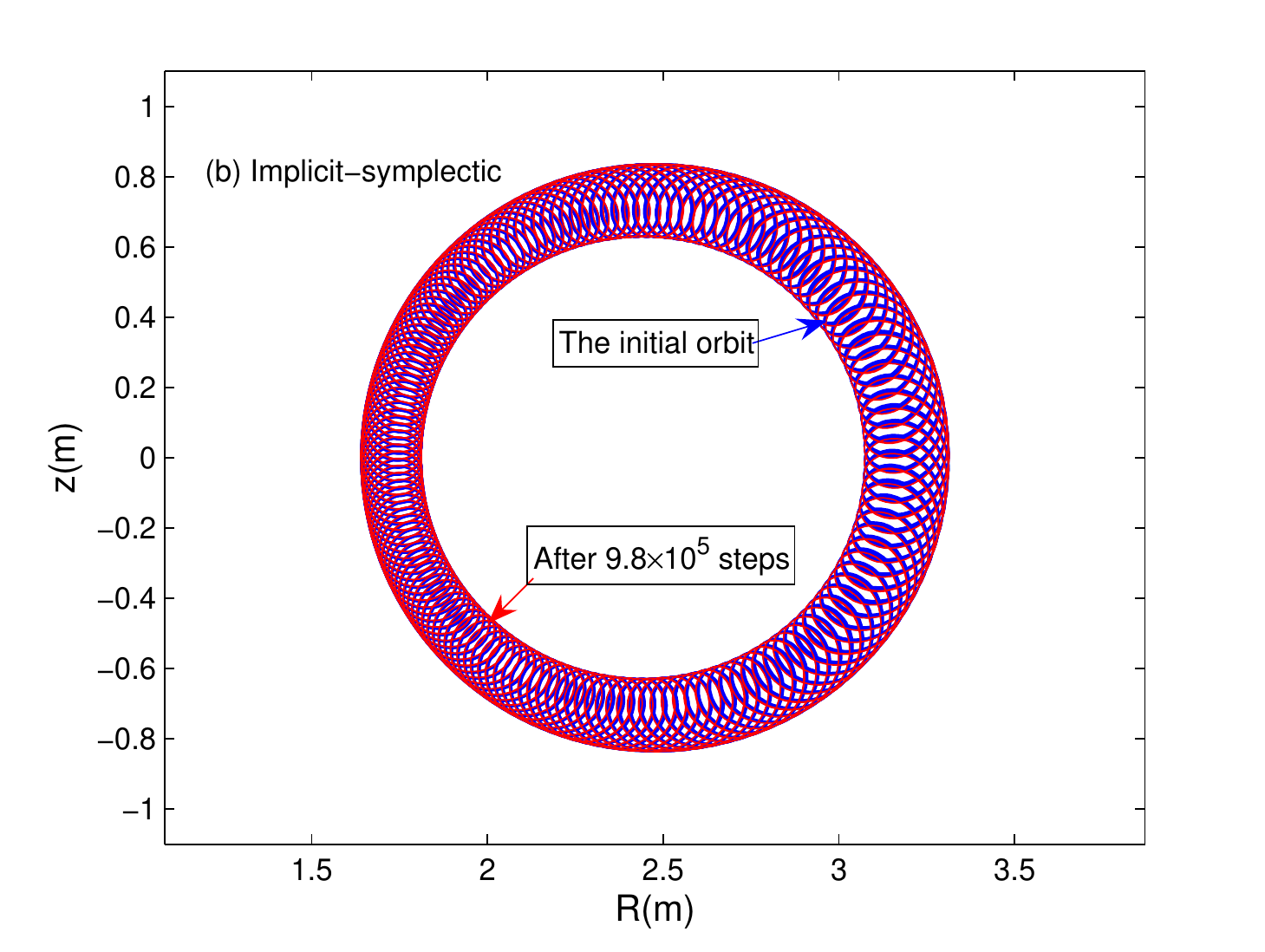}

\includegraphics[scale=0.55]{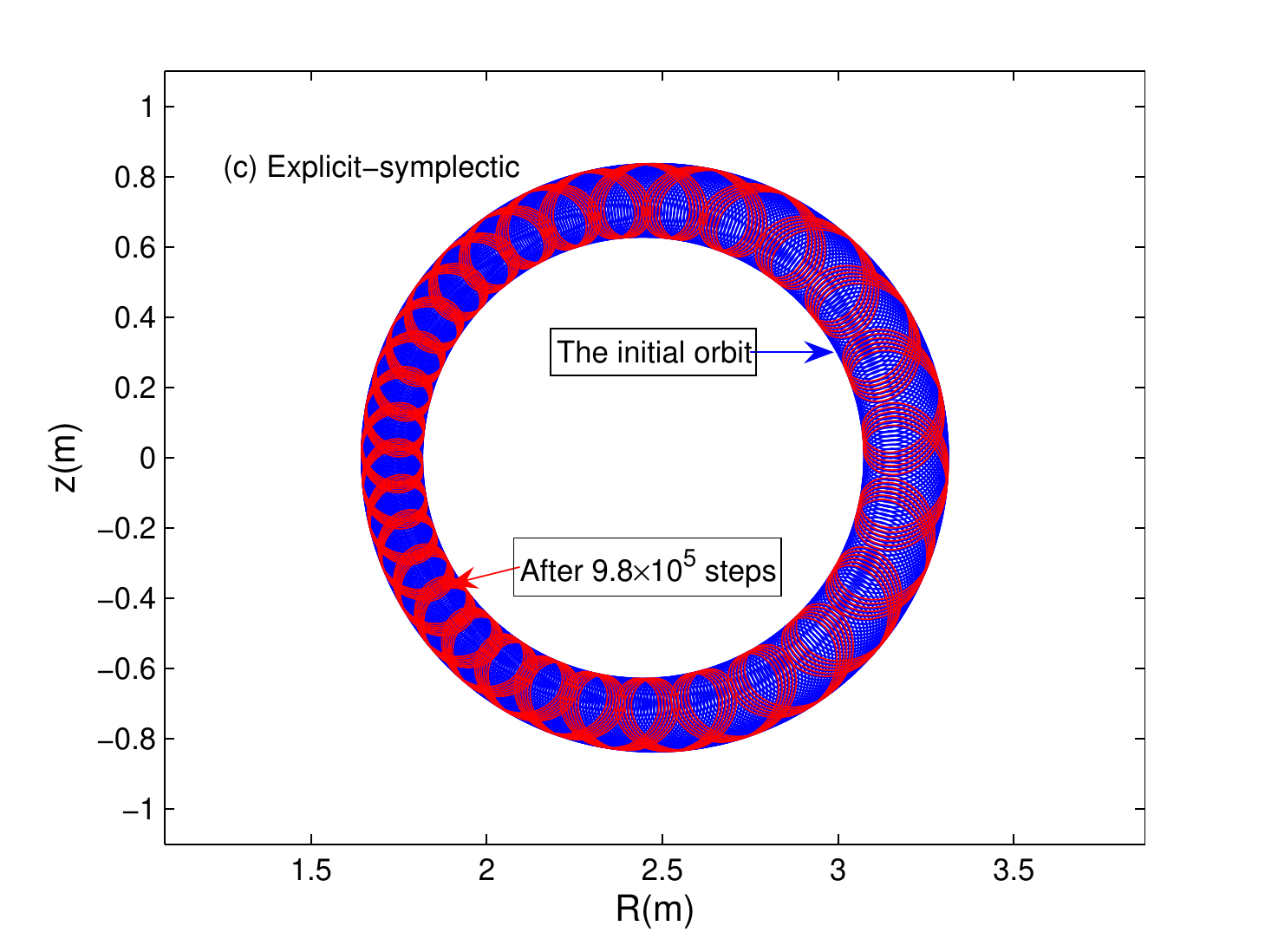}\includegraphics[scale=0.55]{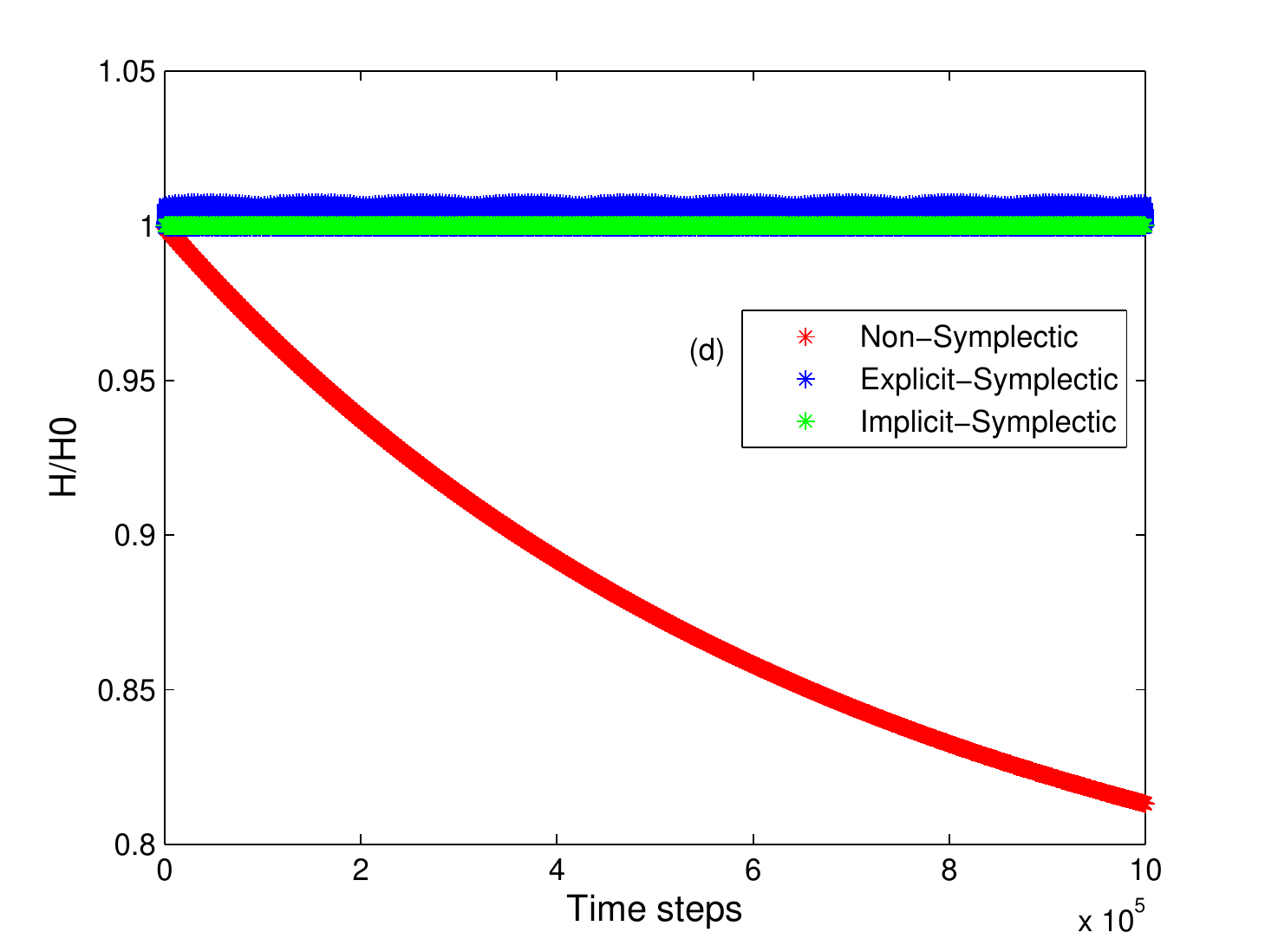}

\protect\caption{Simulations of long-term dynamics of a $3.5$MeV $\alpha-$particle
in a tokamak. The initial orbits are plotted using blue lines, and
the orbits after $9.8\times10^{5}$ steps are plotted using red lines.
(a) Numerical orbit obtained by a non-symplectic RK3 method. (b) The
orbit obtained by the IS2 method. (c) The numerical orbit by the ES2
method $\Psi_{\Delta t}^{2}$. (d) The normalized energy $H/H_{0}$
of three methods are plotted as functions of simulation time step.
\label{fig3}}
\end{figure}

The initial position and velocity of the $\alpha-$particle are $\mathbf{x}_{0}=(3.15,0,0)m$
and $\mathbf{v}_{0}=(0.016,0.04,0)c$, where $c$ is the speed of
light, and the simulation time-step is set to be $\text{\ensuremath{\Delta}}t=0.1\times10^{-8}s$.
Displayed in Fig.$\,$\ref{fig3} is the comparison of transit orbits
calculated by the non-symplectic third order Runge-Kutta (RK3) method,
second order implicit symplectic mid-point (IS2) method and the explicit
second symplectic (ES2) algorithm $\Psi_{\Delta t}^{2}$. It is expected
that the orbit consists of a fast, small scale gyro-motion due to
Lorentz force, and a slow, large scale transit motion induced by the
inhomogeneity of the magnetic field. In Fig. \ref{fig3}, the small
circles of a few centimeters are the fast gyro-motion, and the large
circles about half meter in size in the $RZ-$ plane is the large
scale transit dynamics. Figure.$\,$\ref{fig3}(a) shows that the
orbit obtained by the non-symplectic RK3 method after $9.8\times10^{5}$
time steps is not accurate any more, while the orbits calculated by
the IS2 method in Fig.$\,$\ref{fig3}(b) and ES2 algorithm $\Psi_{\Delta t}^{2}$
in Fig.$\,$\ref{fig3}(c) are accurate for all time steps and form
closed transit orbits. The long-term energy by non-symplectic method
gradually decreases without bound due to numerical errors. On the
contrary, for the symplectic integrators, the energy errors are bounded
by a small number for all time. This fact is clearly demonstrated
in Fig.$\,$\ref{fig3}(d), where normalized energy for the three
algorithms are plotted.

To illustrate the efficiency of the explicit symplectic algorithms
developed, the CPU time used by the three methods for calculating
the charged particle dynamics is listed in Table.~\ref{Tab1}. The
numerical calculation consists of $10^{6}$ time-steps, and is carried
using on a Inter Core $i5-4200U$ CPU. It's clear that the ES2 algorithm
$\Psi_{\Delta t}^{2}$ is much more efficient than the IS2 algorithm.

\begin{table}
\begin{tabular}{|c|c|c|c|}
\hline
 & RK3 & IS2 & ES2\tabularnewline
\hline
\hline
CPU time & 2109s & 3446s & 1212s\tabularnewline
\hline
\end{tabular}\protect\caption{CPU time used by the three algorithms for charged particle dynamics
in a tokamak. }
\label{Tab1}
\end{table}

\section{Conclusion}

In this paper, we have constructed explicit symplectic algorithms
for dynamics of charged particle by combining the familiar sum-split
method with a specially designed generating function method. The newly
developed algorithms are expected to significantly extend the applicability
of symplectic algorithms to physics problems which contain a large
number of degrees of freedom and require accuracy, fidelity and efficiency
of long-term dynamics, such as the classical particle-field system
described by the Vlasov-Maxwell equations \cite{Qin14-043102}.
\begin{acknowledgments}
This research is supported by the National Natural Science Foundation
of China (NSFC-11305171, 11505186, 11575185, 11575186), ITER-China
Program (2015GB111003, 2014GB124005), the Fundamental Research Funds
for the Central Universities (No. WK2030040068), China Postdoctoral
Science Foundation (No. 2015M581994), the CAS Program for Interdisciplinary
Collaboration Team, and the Geo-Algorithmic Plasma Simulator (GAPS)
Project.
\end{acknowledgments}
%

%

\section*{Appendix}

We will prove the explicit algorithm $\Psi_{\Delta t}^{2}$ given
by Eq.\,\eqref{eq:2thes} is a second order method for the Hamiltonian
\begin{equation}
H(\mathbf{x},\mathbf{p})=\dfrac{1}{2}(\mathbf{p}-\mathbf{A}(\mathbf{x}))^{2}+\phi(\mathbf{x})\thinspace.\label{eq:Hnr-1}
\end{equation}
To simplify the notation, we have taken $m=1$ and $q=1.$ There are
three steps in the proof.

Step 1: To prove $\psi_{3}^{\Delta t/2}\circ\psi_{4}^{\Delta t/2}\circ\psi_{5}^{\Delta t}\circ\psi_{4}^{\Delta t/2}\circ\psi_{3}^{\Delta t/2}$
is a numerical method of order 2 for the sub-Hamiltonian system with
Hamiltonian
\begin{equation}
H^{1}(\mathbf{x},\mathbf{p})=-\mathbf{p}_{1}\mathbf{A}_{1}(\mathbf{x})-\mathbf{p}_{2}\mathbf{A}_{2}(\mathbf{x})-\mathbf{p}_{3}\mathbf{A}_{3}(\mathbf{x})\thinspace.
\end{equation}
Since $\psi_{2+i}^{\Delta t},\: i=1,2,3,$ is numerical method of
order 2 for Hamiltonian system generated by $H_{2+i}=-\mathbf{p}_{i}\mathbf{A}_{i}(\mathbf{x})$,
it can be rewritten as
\begin{equation}
\begin{split} & \psi_{2+i}^{\Delta t}(\mathbf{x}^{0},\mathbf{p}^{0})\\
= & \left(I+\Delta t\left(\begin{array}{c}
-\mathbf{A}_{i}\mathbf{e}_{i}\\
\mathbf{p}_{i}\nabla\mathbf{A}_{i}
\end{array}\right)+\dfrac{\Delta t^{2}}{2}\left(\begin{array}{c}
\dfrac{\partial\mathbf{A}_{i}}{\partial\mathbf{x}_{i}}\mathbf{A}_{i}\mathbf{e}_{i}\\
\mathbf{p}_{i}\left(\dfrac{\partial\mathbf{A}_{i}}{\partial\mathbf{x}_{i}}\nabla\mathbf{A}_{i}-\mathbf{A}_{i}\nabla\dfrac{\partial\mathbf{A}_{i}}{\partial\mathbf{x}_{i}}\right)
\end{array}\right)+O(\Delta t^{3})\right)(\mathbf{x}^{0},\mathbf{p}^{0})\thinspace,
\end{split}
\end{equation}
where $\mathbf{e}_{i}$ is the unit vector in the $i$-th Cartesian
direction. The composition method $\psi_{3}^{\Delta t/2}\circ\psi_{4}^{\Delta t/2}\circ\psi_{5}^{\Delta t}\circ\psi_{4}^{\Delta t/2}\circ\psi_{3}^{\Delta t/2}$
can be obtained using the iterations step by step as follows,
\begin{equation}
\begin{split} & \psi_{3}^{\Delta t/2}\circ\psi_{4}^{\Delta t/2}\circ\psi_{5}^{\Delta t}\circ\psi_{4}^{\Delta t/2}\circ\psi_{3}^{\Delta t/2}(\mathbf{x}^{0},\mathbf{p}^{0})\\
= & \left(I+\dfrac{\Delta t}{2}\left(\begin{array}{c}
-\mathbf{A}_{1}\mathbf{e}_{1}\\
\mathbf{p}_{1}\nabla\mathbf{A}_{1}
\end{array}\right)+\dfrac{\Delta t^{2}}{8}\left(\begin{array}{c}
\dfrac{\partial\mathbf{A}_{1}}{\partial\mathbf{x}_{1}}\mathbf{A}_{1}\mathbf{e}_{1}\\
\mathbf{V}_{11}
\end{array}\right)+O(\Delta t^{3})\right)\psi_{4}^{\Delta t/2}\circ\psi_{5}^{\Delta t}\circ\psi_{4}^{\Delta t/2}\circ\psi_{3}^{\Delta t/2}(\mathbf{x}^{0},\mathbf{p}^{0})\\
= & \left(I+\dfrac{\Delta t}{2}\left(\begin{array}{c}
-\mathbf{A}_{1}\mathbf{e}_{1}-\mathbf{A}_{2}\mathbf{e}_{2}\\
\mathbf{p}_{2}\nabla\mathbf{A}_{2}+\mathbf{p}_{1}\nabla\mathbf{A}_{1}
\end{array}\right)+O(\Delta t^{3})\right)\psi_{5}^{\Delta t}\circ\psi_{4}^{\Delta t/2}\circ\psi_{3}^{\Delta t/2}(\mathbf{x}^{0},\mathbf{p}^{0})\\
 & +\dfrac{\Delta t^{2}}{8}\left(\begin{array}{c}
\dfrac{\partial\mathbf{A}_{1}}{\partial\mathbf{x}_{1}}\mathbf{A}_{1}\mathbf{e}_{1}+\dfrac{\partial\mathbf{A}_{2}}{\partial\mathbf{x}_{2}}\mathbf{A}_{2}\mathbf{e}_{2}+2\dfrac{\partial\mathbf{A}_{1}}{\partial\mathbf{x}_{2}}\mathbf{A}_{2}\\
\sum_{i=1}^{2}\mathbf{V}_{ii}+2\mathbf{V}_{21}
\end{array}\right)\psi_{5}^{\Delta t}\circ\psi_{4}^{\Delta t/2}\circ\psi_{3}^{\Delta t/2}(\mathbf{x}^{0},\mathbf{p}^{0})\\
= & \left(I+\dfrac{\Delta t}{2}\left(\begin{array}{c}
-\mathbf{A}_{1}\mathbf{e}_{1}-\mathbf{A}_{2}\mathbf{e}_{2}-2\mathbf{A}_{3}\mathbf{e}_{3}\\
\mathbf{p}_{2}\nabla\mathbf{A}_{2}+\mathbf{p}_{1}\nabla\mathbf{A}_{1}+2\mathbf{p}_{3}\nabla\mathbf{A}_{3}
\end{array}\right)+O(\Delta t^{3})\right)\psi_{4}^{\Delta t/2}\circ\psi_{3}^{\Delta t/2}(\mathbf{x}^{0},\mathbf{p}^{0})\\
 & \dfrac{\Delta t^{2}}{8}\left(\begin{array}{c}
\dfrac{\partial\mathbf{A}_{1}}{\partial\mathbf{x}_{1}}\mathbf{A}_{1}\mathbf{e}_{1}+\dfrac{\partial\mathbf{A}_{2}}{\partial\mathbf{x}_{2}}\mathbf{A}_{2}\mathbf{e}_{2}+2\dfrac{\partial\mathbf{A}_{1}}{\partial\mathbf{x}_{2}}\mathbf{A}_{2}\mathbf{e}_{1}+4\dfrac{\partial\mathbf{A}}{\partial\mathbf{x}_{3}}\mathbf{A}_{3}\\
\sum_{i=1}^{2}\mathbf{V}_{ii}+4\mathbf{V}_{33}+4\mathbf{V}_{32}+4\mathbf{V}_{31}+2\mathbf{V}_{21}
\end{array}\right)\psi_{4}^{\Delta t/2}\circ\psi_{3}^{\Delta t/2}(\mathbf{x}^{0},\mathbf{p}^{0})\\
= & \left(I+\dfrac{\Delta t}{2}\left(\begin{array}{c}
-\mathbf{A}_{1}\mathbf{e}_{1}-2\mathbf{A}_{2}\mathbf{e}_{2}-2\mathbf{A}_{3}\mathbf{e}_{3}\\
2\mathbf{p}_{2}\nabla\mathbf{A}_{2}+\mathbf{p}_{1}\nabla\mathbf{A}_{1}+2\mathbf{p}_{3}\nabla\mathbf{A}_{3}
\end{array}\right)+O(\Delta t^{3})\right)\psi_{3}^{\Delta t/2}(\mathbf{x}^{0},\mathbf{p}^{0})\\
 & +\dfrac{\Delta t^{2}}{8}\left(\begin{array}{c}
\dfrac{\partial\mathbf{A}_{1}}{\partial\mathbf{x}_{1}}\mathbf{A}_{1}\mathbf{e}_{1}+4\dfrac{\partial\mathbf{A}}{\partial\mathbf{x}_{2}}\mathbf{A}_{2}+4\dfrac{\partial\mathbf{A}}{\partial\mathbf{x}_{3}}\mathbf{A}_{3}\\
\mathbf{V}_{11}+4\mathbf{V}_{22}+4\mathbf{V}_{33}+4\mathbf{V}_{32}+4\mathbf{V}_{31}+4\mathbf{V}_{21}+4\mathbf{V}_{23}
\end{array}\right)\psi_{3}^{\Delta t/2}(\mathbf{x}^{0},\mathbf{p}^{0})\\
= & \left(I+\Delta t\left(\begin{array}{c}
-\mathbf{A}\\
\left(\dfrac{\partial\mathbf{A}}{\partial\mathbf{x}}\right)^{T}\mathbf{p}
\end{array}\right)+\dfrac{\Delta t^{2}}{2}\left(\begin{array}{c}
\dfrac{\partial\mathbf{A}}{\partial\mathbf{x}}\mathbf{A}\\
\left(\dfrac{\partial\mathbf{A}}{\partial\mathbf{x}}\right)^{T}\left(\dfrac{\partial\mathbf{A}}{\partial\mathbf{x}}\right)^{T}\mathbf{p}-\left(\dfrac{\partial\mathbf{A}}{\partial\mathbf{x}}\right)_{x}^{T}\mathbf{A}\mathbf{p}
\end{array}\right)+O(\Delta t^{3})\right)(\mathbf{x}^{0},\mathbf{p}^{0})\thinspace,
\end{split}
\end{equation}
where $\mathbf{V}_{ij}=\mathbf{p}_{i}\dfrac{\partial\mathbf{A}_{i}}{\partial\mathbf{x}_{j}}\nabla\mathbf{A}_{j}-\mathbf{p}_{j}\mathbf{A}_{i}\nabla\dfrac{\partial\mathbf{A}_{j}}{\partial\mathbf{x}_{i}}$.
This shows that the $\psi_{3}^{\Delta t/2}\circ\psi_{4}^{\Delta t/2}\circ\psi_{5}^{\Delta t}\circ\psi_{4}^{\Delta t/2}\circ\psi_{3}^{\Delta t/2}$
is of order 2.

Step 2: To prove $\varphi_{2}^{\Delta t/2}\circ\psi_{3}^{\Delta t/2}\circ\psi_{4}^{\Delta t/2}\circ\psi_{5}^{\Delta t}\circ\psi_{4}^{\Delta t/2}\circ\psi_{3}^{\Delta t/2}\circ\varphi_{2}^{\Delta t/2}$
is of order 2 for the sub-Hamiltonian system with Hamiltonian
\begin{equation}
H^{2}(\mathbf{x},\mathbf{p})=-\mathbf{p}_{1}\mathbf{A}_{1}(\mathbf{x})-\mathbf{p}_{2}\mathbf{A}_{2}(\mathbf{x})-\mathbf{p}_{3}\mathbf{A}(\mathbf{x})+\dfrac{1}{2}\mathbf{A}^{2}(\mathbf{x})+\phi(\mathbf{x}).
\end{equation}
 As proved in Step 1, the iteration $\psi_{3}^{\Delta t/2}\circ\psi_{4}^{\Delta t/2}\circ\psi_{5}^{\Delta t}\circ\psi_{4}^{\Delta t/2}\circ\psi_{3}^{\Delta t/2}$
is of order 2, and
\begin{equation}
\varphi_{2}^{\Delta t}(\mathbf{x}^{0},\mathbf{p}^{0})=\left[I-\Delta t\left(\begin{array}{c}
0\\
\left(\dfrac{\partial\mathbf{A}}{\partial\mathbf{x}}\right)^{T}\mathbf{A}+\nabla\phi
\end{array}\right)\right](\mathbf{x}^{0},\mathbf{p}^{0})\thinspace.
\end{equation}
The following calculation shows that $\varphi_{2}^{\Delta t/2}\circ\psi_{3}^{\Delta t/2}\circ\psi_{4}^{\Delta t/2}\circ\psi_{5}^{\Delta t}\circ\psi_{4}^{\Delta t/2}\circ\psi_{3}^{\Delta t/2}\circ\varphi_{2}^{\Delta t/2}$
has accuracy of order 2,
\begin{equation}
\begin{split} & \varphi_{2}^{\Delta t/2}\circ\psi_{3}^{\Delta t/2}\circ\psi_{4}^{\Delta t/2}\circ\psi_{5}^{\Delta t}\circ\psi_{4}^{\Delta t/2}\circ\psi_{3}^{\Delta t/2}\circ\varphi_{2}^{\Delta t/2}(\mathbf{x}^{0},\mathbf{p}^{0})\\
= & \left[I-\dfrac{\Delta t}{2}\left(\begin{array}{c}
0\\
\left(\dfrac{\partial\mathbf{A}}{\partial\mathbf{x}}\right)^{T}\mathbf{A}+\nabla\phi
\end{array}\right)\right]\psi_{3}^{\Delta t/2}\circ\psi_{4}^{\Delta t/2}\circ\psi_{5}^{\Delta t}\circ\psi_{4}^{\Delta t/2}\circ\psi_{3}^{\Delta t/2}\circ\varphi_{2}^{\Delta t/2}(\mathbf{x}^{0},\mathbf{p}^{0})\\
= & \left[I+\Delta t\left(\begin{array}{c}
-\mathbf{A}\\
\left(\dfrac{\partial\mathbf{A}}{\partial\mathbf{x}}\right)^{T}\mathbf{p}-\dfrac{1}{2}\left(\left(\dfrac{\partial\mathbf{A}}{\partial\mathbf{x}}\right)^{T}\mathbf{A}+\nabla\phi\right)
\end{array}\right)+O(\Delta t^{3})\right]\varphi_{2}^{\Delta t/2}(\mathbf{x}^{0},\mathbf{p}^{0})\\
 & +\dfrac{\Delta t^{2}}{2}\left(\begin{array}{c}
\dfrac{\partial\mathbf{A}}{\partial\mathbf{x}}\mathbf{A}\\
\left(\dfrac{\partial\mathbf{A}}{\partial\mathbf{x}}\right)^{T}\left(\dfrac{\partial\mathbf{A}}{\partial\mathbf{x}}\right)^{T}\mathbf{p}-\left(\dfrac{\partial\mathbf{A}}{\partial\mathbf{x}}\right)_{x}^{T}\mathbf{A}\mathbf{p}+\nabla_{xx}(\dfrac{\mathbf{A}^{2}}{2})\mathbf{A}+\nabla_{xx}\phi\mathbf{A}
\end{array}\right)\varphi_{2}^{\Delta t/2}(\mathbf{x}^{0},\mathbf{p}^{0})\\
= & \left[I+\Delta t\left(\begin{array}{c}
-\mathbf{A}\\
\left(\dfrac{\partial\mathbf{A}}{\partial\mathbf{x}}\right)^{T}\mathbf{p}-\left(\dfrac{\partial\mathbf{A}}{\partial\mathbf{x}}\right)^{T}\mathbf{A}-\nabla\phi
\end{array}\right)+O(\Delta t^{3})\right](\mathbf{x}^{0},\mathbf{p}^{0})\\
 & +\dfrac{\Delta t^{2}}{2}\left(\begin{array}{c}
\dfrac{\partial\mathbf{A}}{\partial\mathbf{x}}\mathbf{A}\\
\left(\dfrac{\partial\mathbf{A}}{\partial\mathbf{x}}\right)^{T}\left(\dfrac{\partial\mathbf{A}}{\partial\mathbf{x}}\right)^{T}\mathbf{p}-\left(\dfrac{\partial\mathbf{A}}{\partial\mathbf{x}}\right)_{x}^{T}\mathbf{A}\mathbf{p}+\nabla_{xx}(\dfrac{\mathbf{A}^{2}}{2})\mathbf{A}+\nabla_{xx}\phi\mathbf{A}
\end{array}\right)(\mathbf{x}^{0},\mathbf{p}^{0})\\
 & +\dfrac{\Delta t^{2}}{2}\left(\begin{array}{c}
0\\
-\left(\dfrac{\partial\mathbf{A}}{\partial\mathbf{x}}\right)^{T}\left(\left(\dfrac{\partial\mathbf{A}}{\partial\mathbf{x}}\right)^{T}\mathbf{A}+\nabla\phi\right)
\end{array}\right)(\mathbf{x}^{0},\mathbf{p}^{0})\thinspace.
\end{split}
\end{equation}

Step 3: To prove $\varphi_{1}^{\Delta t/2}\circ\varphi_{2}^{\Delta t/2}\circ\psi_{3}^{\Delta t/2}\circ\psi_{4}^{\Delta t/2}\circ\psi_{5}^{\Delta t}\circ\psi_{4}^{\Delta t/2}\circ\psi_{3}^{\Delta t/2}\circ\varphi_{2}^{\Delta t/2}\circ\varphi_{1}^{\Delta t/2}$
is of order 2 for the Hamiltonian Eq.\,\eqref{eq:Hnr-1}. The iteration
$\varphi_{1}^{\Delta t}$ is
\begin{equation}
\varphi_{1}^{\Delta t}(\mathbf{x}^{0},\mathbf{p}^{0})=\left(I+\Delta t\left(\begin{array}{c}
\mathbf{p}\\
0
\end{array}\right)\right)(\mathbf{x}^{0},\mathbf{p}^{0})\thinspace.
\end{equation}
Combining with the second order iteration $\varphi_{2}^{\Delta t/2}\circ\psi_{3}^{\Delta t/2}\circ\psi_{4}^{\Delta t/2}\circ\psi_{5}^{\Delta t}\circ\psi_{4}^{\Delta t/2}\circ\psi_{3}^{\Delta t/2}\circ\varphi_{2}^{\Delta t/2}$
proved in Step 2, we obtain
\begin{equation}
\begin{split}\Psi_{\Delta t}^{2} & =\varphi_{1}^{\Delta t/2}\circ\varphi_{2}^{\Delta t/2}\circ\psi_{3}^{\Delta t/2}\circ\psi_{4}^{\Delta t/2}\circ\psi_{5}^{\Delta t}\circ\psi_{4}^{\Delta t/2}\circ\psi_{3}^{\Delta t/2}\circ\varphi_{2}^{\Delta t/2}\circ\varphi_{1}^{\Delta t/2}\\
 & =\left(I+\Delta t\left(\begin{array}{c}
\mathbf{p}\\
0
\end{array}\right)\right)\varphi_{2}^{\Delta t/2}\circ\psi_{3}^{\Delta t/2}\circ\psi_{4}^{\Delta t/2}\circ\psi_{5}^{\Delta t}\circ\psi_{4}^{\Delta t/2}\circ\psi_{3}^{\Delta t/2}\circ\varphi_{2}^{\Delta t/2}\circ\varphi_{1}^{\Delta t/2}(\mathbf{x}^{0},\mathbf{p}^{0})\\
 & =\left(I+\Delta t\left(\begin{array}{c}
\mathbf{p-}\mathbf{A}\\
\left(\dfrac{\partial\mathbf{A}}{\partial\mathbf{x}}\right)^{T}\mathbf{p}-\left(\dfrac{\partial\mathbf{A}}{\partial\mathbf{x}}\right)^{T}\mathbf{A}-\nabla\phi
\end{array}\right)+O(\Delta t^{3})\right)(\mathbf{x}^{0},\mathbf{p}^{0})\\
 & +\dfrac{\Delta t^{2}}{2}\left(\begin{array}{c}
\left(\left(\dfrac{\partial\mathbf{A}}{\partial\mathbf{x}}\right)^{T}-\dfrac{\partial\mathbf{A}}{\partial\mathbf{x}}\right)\left(\mathbf{p}-\mathbf{A}\right)-\nabla\phi\\
\left(\dfrac{\partial\mathbf{A}}{\partial\mathbf{x}}\right)^{T}\left(\dfrac{\partial\mathbf{A}}{\partial\mathbf{x}}\right)^{T}\mathbf{p}+\left(\dfrac{\partial\mathbf{A}}{\partial\mathbf{x}}\right)_{x}^{T}(\mathbf{p}-\mathbf{A})\mathbf{p}
\end{array}\right)(\mathbf{x}^{0},\mathbf{p}^{0})\\
 & +\dfrac{\Delta t^{2}}{2}\left(\begin{array}{c}
0\\
-\nabla_{xx}(\dfrac{\mathbf{A}^{2}}{2}+\phi)(\mathbf{p}-\mathbf{A})-\left(\dfrac{\partial\mathbf{A}}{\partial\mathbf{x}}\right)^{T}\left(\left(\dfrac{\partial\mathbf{A}}{\partial\mathbf{x}}\right)^{T}\mathbf{A}+\nabla\phi\right)
\end{array}\right)(\mathbf{x}^{0},\mathbf{p}^{0})\thinspace,
\end{split}
\end{equation}
which shows that $\Psi_{\Delta t}^{2}$ is a second order method.
\end{document}